\documentclass[amsmath,amssymb,reprint]{revtex4-1}
\usepackage{graphicx,epstopdf}
\usepackage{natbib}

\begin{document}

\title{Optical quantum memory for ultrafast photons using molecular alignment}

\author{G.S. Thekkadath, K. Heshami, D.G. England, P.J. Bustard, B.J. Sussman, 
and M. Spanner$^{\ast}$\thanks{$^\ast$Corresponding author. Email: michael.spanner@nrc.ca}}

\affiliation{National Research Council of Canada, 100 Sussex Drive, Ottawa, Ontario, K1A 0R6, Canada}

\begin{abstract}
The absorption of broadband photons in atomic ensembles requires either an
effective broadening of the atomic transition linewidth, or an off-resonance
Raman interaction.  Here we propose a scheme for a quantum memory capable of
storing and retrieving ultrafast photons in an ensemble of two-level atoms by
using a propagation medium with a time-dependent refractive index generated
from aligning an ensemble of gas-phase diatomic molecules. The refractive index
dynamics generates an effective longitudinal inhomogeneous broadening of the
two-level transition. We numerically demonstrate this scheme for
storage and retrieval of a weak pulse as short as 50\,fs, with a storage time
of up to 20\,ps. With additional optical control of the molecular alignment
dynamics, the storage time can be extended about one nanosecond leading to
time-bandwidth products of order $10^4$.  This scheme could in principle be
achieved using either a hollow-core fiber or a high-pressure gas cell, in a
gaseous host medium comprised of diatomic molecules and a two-level atomic
vapor at room temperature.
\end{abstract}

\maketitle

\section{Introduction}
Controlling light-matter interfaces is essential to further develop optical
quantum information processing elements. Optical quantum memories are one of
the primary goals of this development, and their
applications~\cite{heshami2016,bussieres2013} are likely to extend beyond their role in
quantum repeaters~\cite{sangouard2011quantum}. Many recent efforts have been
devoted to their implementation in atomic ensembles~\cite{hammerer2010quantum},
and in particular, memories compatible with ultrafast light are desirable for
their large time-bandwidth products and fast operational
speeds~\cite{england2013from,england2015storage,bustard2013toward}.

Absorbing broadband photons in atomic ensembles requires either a controlled
inhomogeneous broadening of the atomic transition or an off-resonant
Raman interaction, such as Raman memories~\cite{reim2010towards}, atomic
frequency combs~\cite{saglamyurek2011broadband}, and gradient echo memories
(GEMs)~\cite{hetet2008electro}. This latter approach uses a longitudinal
controlled reversible inhomogeneous broadening of the transition frequency of
the medium to control the photon storage and
retrieval~\cite{moiseev2001complete}. The inhomogeneous broadening is induced
by a spatial gradient in an external static field and modifies the transition
frequency of the atoms. If the broadening is reversed, the time-reversal
symmetry of the Maxwell-Bloch equations predicts that the absorbed light is
subsequently re-emitted~\cite{kraus2006quantum}. GEMs have been realized with a
spatially-dependent linear Stark or Zeeman shift in rare-earth-doped crystals
and warm atomic
vapors~\cite{alexander2006photon,hetet2008electro,hosseini2009coherent,hedges2010efficient}.
The strength and reversal timescale of the external field impose technical and
fundamental limitations on the operational bandwidth of the GEM, and
consequently the experimental demonstrations have remained limited to
microsecond-duration pulses.

Ref.~\cite{clark2012photonic} showed that having a propagation medium with a
linear time-dependent refractive index is equivalent to the longitudinal
(spatial) inhomogeneous broadening in a GEM. The time-dependent refractive
index can be generated in a host crystal exposed to a varying electric field,
and induces a linear spatially-dependent frequency shift in the propagating
signal pulse that mimics the effect of a varying resonance atomic frequency. This scheme relies on the temporal modulation of an
external electric field which cannot be switched quickly enough to store
ultrafast photons.

In this paper, we employ molecular alignment techniques to engineer the
ultrafast refractive index dynamics necessary for storing femtosecond pulses in
an atomic ensemble. Molecular alignment is a well-studied
phenomenon~\cite{stapelfeldt2003colloquium} where a non-resonant ultrafast
laser pulse impulsively generates a coherent excitation in the rotational
levels of an ensemble of molecules via the dynamic Stark
effect~\cite{sussman2011five}. It has been used to shape light
pulses~\cite{PhysRevLett.88.013903,PhysRevLett.88.203901,PhysRevLett.104.193902},
dissociate diatomic molecules~\cite{villeneuve2000forced}, in optical imaging~\cite{wu2010optical}, and for quantum
control
scenarios~\cite{shapiro2003strong,spanner2004coherent,PhysRevLett.93.233601}.
We propose to use such an interaction in a gaseous host medium containing
diatomic molecules and two-level atoms (\textit{e.g.} warm atomic vapor) to
optically prepare refractive index dynamics which can be used for a GEM-type
storage and retrieval of ultrafast photons.  As a proof-of-principle, we
numerically simulate this interaction for the case of a mixed CO$_2$/$^{87}$Rb
gas, and store the signal using the D$_1$ transition (795\,nm) of $^{87}$Rb
atoms~\cite{steck2008rubidium}. 

The paper is organized as follows: in Sec.~\ref{sec:theory}, we briefly explain
the proposed scheme, and present the necessary molecular alignment and
Maxwell-Bloch equations to numerically simulate the process of storing and
retrieving a signal pulse. In Sec.~\ref{sec:discussion}, we discuss the results of
the simulation and propose a possible experimental implementation.  Finally in
Sec.~\ref{sec:conclusion}, we conclude the paper.

\section{Theory}
\label{sec:theory}

\subsection{General Concept}
\begin{figure}
    \centering
    \includegraphics[width=\columnwidth]{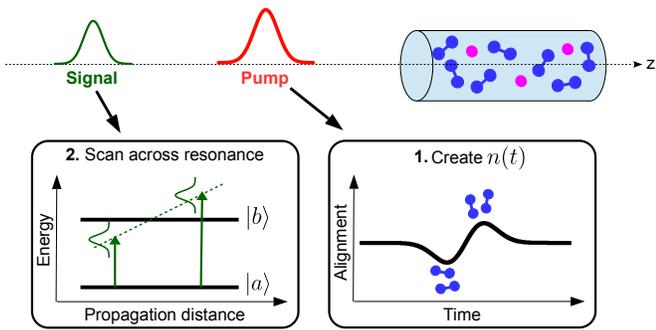}
	\caption{Input pulse sequence for the memory scheme.  A pump pulse and a
delayed signal pulse are sent into a mixed gas medium comprised of CO$_2$
molecules (blue) and $^{87}$Rb atoms (pink).  Step 1: The pump generates
time-dependent molecular alignment, which causes a corresponding time-dependent
refractive index $n(t)$ in the medium.  Step 2: The delayed signal pulse
experiences the $n(t)$ as it propagates through the medium.  For specific
delays, the $n(t)$ causes the bandwidth of the signal to be scanned across the
atomic resonance and thus absorbed in the two-level atoms.  Step 3: (not shown)
At a later time, the $n(t)$ is reversed due to the natural rotational dynamics
of the aligned molecules, and this reversal of $n(t)$ causes the re-emission of
the signal.}
    \label{fig:scheme}
\end{figure}

The proposed scheme is as follows (see Fig.~\ref{fig:scheme}): (i)
(\textit{pump}) A strong non-resonant pump pulse aligns an ensemble of diatomic
gas molecules and generates a time-dependent refractive index $n(t)$ in the
propagation medium. (ii) (\textit{absorption}) By choosing an appropriate
pump-signal delay, the weak signal pulse propagates in a region in time where
the refractive index is approximately linear, \textit{i.e.} $n(t) \approx n_0 +
\dot{n}t$. This evenly shifts the frequency distribution of the signal pulse by
an amount at least equal to its bandwidth (for a sufficient $\dot{n}$), thereby
scanning all its frequency components over the transition frequency of
two-level atoms such that the entire pulse can be absorbed. (iii)
(\textit{emission}) Later in time, oscillations in the rotational dynamics of
the molecules generate the opposite slope in the refractive index of the
propagation medium, \textit{i.e.} $n(t+\tau)\approx n_0 - \dot{n}t$. This in
turn causes constructive interference in the collective wave function of the
two-level atoms, and results in re-emission of the signal with a delay $\tau$
that defines the storage time of the memory.      

\subsection{Molecular alignment in the host gas: refractive index dynamics
preparation} A non-resonant ultrafast laser pump pulse excites a rotational
wave packet $|\Psi(t)\rangle = \sum_{J,M} A_{J,M}(t)|J,M\rangle$ in a
rigid-rotor type molecule, where $|J,M\rangle$ are orbital angular momentum
eigenstates $\langle \theta,\phi | J,M\rangle = Y_{JM}(\theta,\phi)$, and
$A_{J,M}(t)$ are the coefficients of the rotational excitations. The time
evolution of the rotational wave packet is given by the Schr\"odinger
equation~\cite{friedrich1995alignment}: 
\begin{equation} \label{eqn:tdse}
i\frac{\partial}{\partial t} |\Psi(t)\rangle = [B_0\mathbf{J}^2 -
U_0(t)\cos^2{\theta}] |\Psi(t)\rangle 
\end{equation} 
where $B_0$ is the
rotational constant of the molecule, $\mathbf{J}^2$ is the rotational energy
operator, $U_0(t)\cos^2{\theta}$ describes the interaction between the
(linearly polarized) electric field of the laser and polarizability of the
molecule, and $\theta$ is the relative angle between the laser polarization
vector and the molecular axis. Note that all equations are written in atomic
units throughout the paper, unless otherwise specified. The magnitude of the
laser interaction term is given by 
\begin{equation} 
	U_0(t) = \frac{1}{4}\Delta\alpha\mathcal{E}_0^2\sin^2{\left (\frac{\pi
t}{2\sigma_p}\right )} 
\end{equation} 
where $\mathcal{E}_0$ is the pump field
strength, $\sigma_p$ is the pump pulse duration, and $\Delta\alpha =
\alpha_\parallel - \alpha_\perp$ is the difference in the polarizability of the
molecule along its two axes. 

The alignment generated by this rotational wave packet is characterized by the
expectation value $\langle\cos^2{\theta}\rangle(t)=\langle\psi(t) |
\cos^2{\theta} | \psi(t)\rangle$, and must be averaged over the thermal
Boltzmann distribution of the molecules' initial angular momentum states at a
temperature $T$.  Specifically, each initial $|J,M\rangle$ state present in the
initial thermal distribution is excited by the pump field and generates a
particular $\langle\cos^2{\theta}\rangle_{J,M}(t)$, where the added subscript
denotes that this contribution came from the initial thermally-populated
$|J,M\rangle$ state.  These $\langle\cos^2{\theta}\rangle_{J,M}(t)$ are then further
averaged over the initial Boltzmann distribution to give the thermally-averaged
$\langle\cos^2{\theta}\rangle_T(t)$ 
\begin{equation}
	\langle\cos^2{\theta}\rangle_T(t) = \frac{\sum_{J,M} g_J
	e^{-E_J/k_BT}\langle\cos^2{\theta}\rangle_{J,M}(t)}{\sum_J
	g_J(2J+1)e^{-E_J/k_BT}} 
\end{equation} 
where the rotational energies are $E_J =
B_0J(J+1)$, and $g_J$ is a weighting parameter taking into account nuclear spin
statistics. For an ensemble of molecules in thermal equilibrium, the value
$\langle\cos^2\theta\rangle_T = 1/3$ corresponds to a randomly aligned ensemble
whereas $\langle\cos^2\theta\rangle_T = 1$ ($\langle\cos^2\theta\rangle_T = 0$)
corresponds to an ensemble fully parallel (perpendicular) to the electric field
of the pump.  

\subsection{The Maxwell-Bloch equations}
We use a semi-classical treatment to propagate a weak signal pulse in a medium
consisting of an ensemble of two-level atoms and a time-dependent molecular
alignment. We define a signal pulse with a Gaussian temporal envelope and
central frequency $\omega_0$ as 
\begin{equation}\label{pulse}
E_s(t) = E_0e^{-4ln(2)(t-t_0)^2/\sigma_s^2}e^{-i \omega_0 t} 
\end{equation}
where $\sigma_s$ is the signal pulse duration at the full-width-half-max.  
The two-level atoms with energy
spacing $\omega_{ba}$ and transition dipole matrix element $\mu_{ba}$ interact
with the propagating signal. The resulting dynamics can be described via the
Bloch equations~\cite{boyd2003nonlinear}:
\begin{equation}
\label{eqn:bloch}
\begin{aligned}
\dot{\rho}_{ba}(t) &= -(iw_{ba} + \frac{1}{T_2})\rho_{ba}(t) - i\mu_{ba}E_s(t)\rho_d(t) \\
\dot{\rho}_d(t) &= \frac{-\rho_d(t) + 1}{T_1} + 2i(\mu_{ba}E_s(t)\rho_{ab} - \mu_{ab}E_s(t)\rho_{ba})
\end{aligned}
\end{equation}
where $\rho_d$ is the difference in the excited and ground state populations.
The parameters $T_1$ and $T_2$ take into account the lifetime of the transition
and dipole dephasing, respectively. Eq.~\ref{eqn:bloch} couples to Maxwell's
wave equation (see below) via the total polarization of the medium $P(t)=P_m(t)
+ P_a(t)$, which includes effects due to the rotating molecules and the
two-level atoms, respectively, \textit{i.e.}
\begin{equation}
\begin{aligned}
P_m(t) &= N_m[\alpha_\perp + \Delta\alpha\langle\cos^2{\theta}\rangle_T(t)]E_s(t) \\
P_a(t) &= N_a[\mu_{ab}\rho_{ba}(t)]
\end{aligned}
\end{equation}
where $N_m$ and $N_a$ are the respective number densities. We assume a
spatially-uniform distribution of both molecules and two-level atoms along the
propagation axis. 

The molecular polarization term generates a time-dependent refractive index:
\begin{equation} \label{EqNofT}
n(t) = 1 + 2\pi N_m[\alpha_\perp + \Delta\alpha\langle\cos^2{\theta}\rangle_T(t)].
\end{equation}
which is valid for $N_m \ll 1$. Ignoring transverse dynamics, we use a
simplified Maxwell equation
\begin{equation}
\frac{\partial E(t)}{\partial z} + \frac{1}{c}\frac{\partial E(t)}{\partial t} = -\frac{2\pi}{c}\frac{\partial P(t)}{\partial t}
\label{eqn:maxwell}
\end{equation}
to solve for forward propagating (along $z$) waves
~\cite{couairon2011practitioner, bullough1979solitons}.  We note that this
equation is applicable to the case of a single photon.  Eq.~\ref{eqn:maxwell}
is solved in the moving reference frame of the pump pulse, \textit{i.e.} $\tau
\rightarrow t - \frac{z}{v_s}$ where 
\begin{equation}
v_s = \frac{c}{n_0} = \frac{c}{1+2\pi N_m(\alpha_\perp + \Delta\alpha/3)},
\end{equation}
such that the dynamics of the alignment always begin at $\tau=0$ in the
reference frame of the pump. This assumes the pump is traveling through an
ensemble of initially randomly aligned molecules with
$\langle\cos^2{\theta}\rangle_T = 1/3$. Effects of the molecular alignment on
the pump pulse itself do not strongly affect the generated rotational wave
packet ~\cite{spanner2003optimal}, hence we ignore any modifications to the
pump pulse during propagation.  In the current model, we neglect group velocity
dispersion.  For a specific implementation, care may need to be taken to manage
its effects.  We note, however, that in early studies of pulse compression
using a similar setup, the effects of dispersion were not detrimental
\cite{PhysRevLett.88.013903,kalosha2002generation,spanner2003optimal}. 

\section{Numerical Results and Discussion}
\label{sec:discussion}
\subsection{Basic storage protocol}

\begin{figure}
    \centering
    \includegraphics[width=\columnwidth]{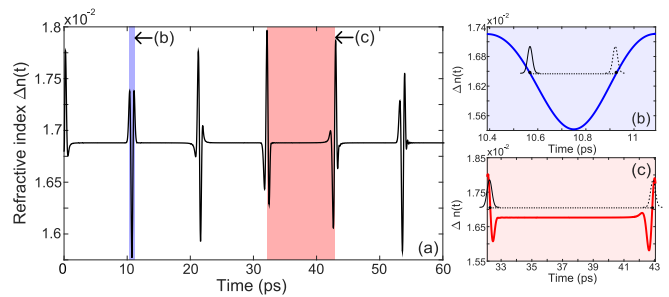}
	\caption{(a) The alignment of an ensemble of CO$_2$ molecules at room
temperature using a strong non-resonant 50\,fs pump pulse generates a
time-dependent refractive index $n(t)$ in the propagation medium. Here we plot
$\Delta n(t)\equiv n(t) - 1$. The molecules continue to rotate well after the
pump pulse is off due to revival dynamics of the generated rotational wave
packet. The highlighted regions (shown in (b) and (c)) provide a suitable
refractive index modulation to achieve absorption and re-emission of a
broadband photon. The memory in Fig.~\ref{fig:O2_memory} is achieved using the
refractive index in (b).}
    \label{fig:alignment}
\end{figure}

Fig.~\ref{fig:alignment}a shows the time-dependent refractive index generated
by aligning an ensemble of CO$_2$ gas molecules 
($\alpha_\perp = 1.97$\,\AA$^3$,
$\Delta \alpha = 2.04$\,\AA$^3$~\cite{miller1990calculation}, $N_m = 1 \times
10^{21}$\,cm$^{-3}$) at room temperature (295\,K) with a strong ($5 \times
10^{13}$\,W/cm$^2$) non-resonant 50\,fs pump pulse.  Here we plot the
refractive index as the deviation from unity $\Delta n(t)\equiv n(t) - 1$. The
bandwidth of the pump should not overlap with the two-level atomic transition
at 795\,nm, which could be readily achieved by using a wavelength in the
1\,-\,2\,$\mu$m regime.  We note that our chosen density $N_m$ represents a
pressure of $\sim$40\,bar which is achievable in a high-pressure gas cell.  The
oscillations seen in Fig.~\ref{fig:alignment}a, which are called rotational
revivals \cite{PhysRevLett.83.4971,stapelfeldt2003colloquium}, reflect the
rotational alignment dynamics of the molecules: The peaks (dips) in $n(t)$
correspond to moments in time where the molecules are preferentially aligned
parallel (perpendicular) to the polarization direction of the pump.  By
changing the delay between the pump and signal pulses, we select a region
within $n(t)$ with the desired shape, as shown in Fig.~\ref{fig:alignment}b and
Fig.~\ref{fig:alignment}c.  Although we report results for the refractive index
in the former region only, any refractive index dynamic containing two opposite
linear slopes separated in time, such as the latter region, is also suitable
for the scheme.

\begin{figure}
    \centering
    \includegraphics[width=\columnwidth]{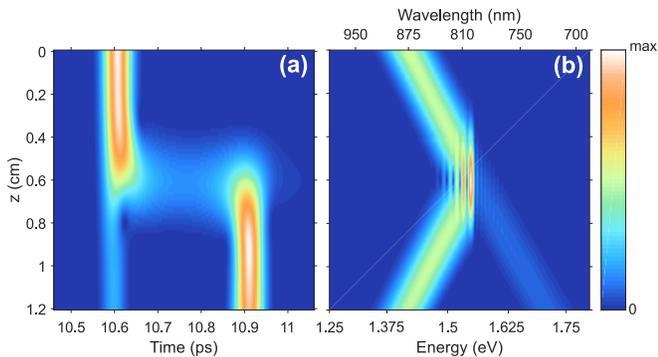}
	\caption{Quantum memory for a weak 50-fs pulse. In (a), the amplitude
$|E(t)|$ of the signal is shown propagating in the reference frame of the pump
in a medium containing two-level atoms and molecules generating the
time-dependent refractive index $n(t)$ (shown in Fig.~\ref{fig:alignment}b).
This causes the signal to be frequency-shifted, as shown in (b): As the signal
propagates, its frequency distribution is scanned over the transition
wavelength (795\,nm) of the two-level atoms such that the entire broadband
pulse is absorbed. The linear slope of the refractive index eventually switches
and the pulse is re-emitted from the medium, and again frequency-shifted, but
in the opposite direction.  Thus the dynamics of the refractive index cause the
signal to be absorbed and then re-emitted from the medium with a delay $\tau$
that defines the storage time of the memory.}
    \label{fig:O2_memory}
\end{figure}

Fig.~\ref{fig:O2_memory}a shows the E-field amplitude $|E(t)|$ of a weak 50-fs
pulse ($\omega_0 = 1.4$\,eV) propagating in the reference frame of the pump. We
set the field amplitude $E_0$ in Eq.~\ref{pulse} to be the amplitude of a
single photon occupying a mode volume defined by interaction distance and the
pulse spot size.  However, we note that the specific quantitative value used
for the magnitude of $E_0$ is not crucial as long as it is in the linear regime
(\textit{i.e.} negligible atomic excitation).  

The linear time-dependent refractive index in the medium causes a
spatially-dependent frequency shift (see Fig.~\ref{fig:O2_memory}b). The
propagation distance required for the signal to be shifted by its entire
bandwidth is determined by the slope of $n(t)$.  For our parameters, we require
a propagation distance of $\sim$1\,cm.  Because the pulse propagates in a
region where the refractive index is linear, the Gaussian shape of the
frequency distribution is preserved as the pulse is frequency-shifted. As the
signal is scanned over the transition wavelength of the two-level atoms
(795\,nm), it is completely absorbed for sufficiently high optical depths (see
Fig.~\ref{fig:efficiency}, 
here we use $N_a = 1.35 \times 10^{16}$\,cm$^{-3}$ and $\mu_{ab}$ = 2.99 $ea_0$).
We assume the two-level atoms have a negligible linewidth relative to the
timescale of refractive index dynamics, and hence ignore relaxation processes
($T_1 = T_2 \rightarrow \infty$ in Eq.~\ref{eqn:bloch}).

After some delay, which we here choose to label as $\tau$, the rotational
dynamics of the molecules causes the refractive index gradient to switch (see
Fig.~\ref{fig:alignment}b), and the signal is re-emitted from the medium, as shown in
Fig.~\ref{fig:O2_memory}a. Note that the opposite gradient in $n(t)$ also
causes the emitted light to be frequency-shifted in the opposite direction,
towards its initial distribution. The storage time is set by $\tau$, and in
this simulation $\tau \approx $ 300\,fs. 
The fringes in Fig.~\ref{fig:O2_memory}b are due to the fact that the
absorbed and emitted pulses are two identical Gaussian functions separated by a
time delay $\tau$, which in Fourier space transforms to a phase term $1 +
e^{i\tau \omega}$, and thus yields fringes whose spacing is given by the
inverse of the storage time of the memory.

The timescale of the linear region of the refractive index oscillation in
Fig.~\ref{fig:alignment}b sets the maximum length of the signal pulse, which in
the case of CO$_2$ is roughly 150\,fs. We note that this is not a limitation of
the scheme as molecular alignment can be achieved with many different
molecules: oscillations in $n(t)$ depend on the moment of inertia of the
molecule being aligned. For instance, a heavier molecule such as Br$_2$ will rotate 
more slowly and hence the refractive index it
generates will also have slower oscillations.  One could also consider using
adiabatic alignment \cite{stapelfeldt2003colloquium} as an alternative to
impulsive alignment since the former operates with picosecond pump pulses. 

Optical control of the molecular alignment process provides additional
versatility to this scheme. For example, undesired dynamics in the alignment,
such as the small oscillation at 32.5\,ps in Fig.~\ref{fig:alignment}c, can be
minimized by temporal shaping of the pump pulse. Although we numerically
verified that the slope in this feature is insufficient to cause re-emission of
the pulse (not shown), it does affects the specific re-emission time at the
subsequent revival at $\sim$43\,ps.

\subsection{Variable read times}

\begin{figure}
    \centering
    \includegraphics[width=\columnwidth]{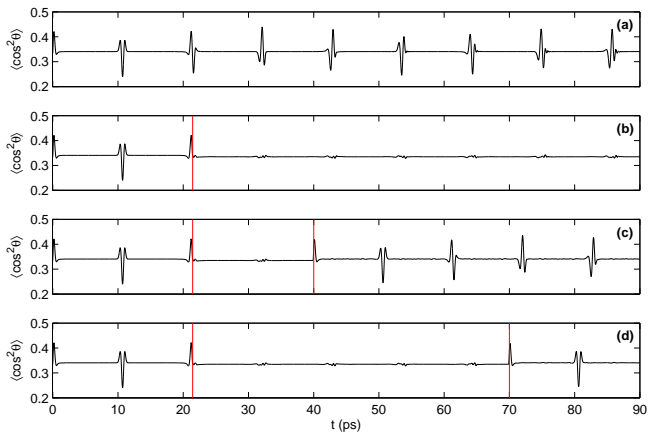}
	\caption{Controlling the rotational revivals. The timings of the control pulses
used in these examples are denoted by the vertical lines.  (a) Initial alignment revivals
generating by the first pump pulse (same revivals as seen in Fig.~2a).
(b) Revivals are attenuated using a control pulse positioned at 21.3 ps .
(c) Attenuated revivals are regenerated using a final control pulse at 40 ps.
(d) Attenuated revivals are regenerated using a final control pulse at 70 ps.  }
    \label{fig:control_revivals}
\end{figure}

The memory scheme we have presented above effectively operates as a delay line
with a fixed storage time.   However, it can be extended to support variable
storage times.  It is possible to attenuate and regenerate the rotational
revivals by using additional control pulses
~\cite{spanner2004coherent,PhysRevLett.93.233601}.  Control over the
alignment revival dynamics is shown in Fig.~\ref{fig:control_revivals}.  Here,
the initial alignment revivals generated by the pump pulse (seen in
Fig.~\ref{fig:control_revivals}, same revivals as shown in Fig.~2(a)) are
attenuated using a control pulse positioned at 21.45 ps (b)-(d).  Recall that
$n(t)$ is proportional to $\langle\cos^2\theta\rangle$ as shown in
Eq.~(\ref{EqNofT}).  The attenuating control pulse is identical to the initial
pump pulse.  This control pulse applies a torque to the molecules just as they
are rotating toward the equatorial plane, and stops the rotations
generated by the first pump pulse.  The rotational distribution of angular momentum states is
then effectively returned to the initial thermal distribution.  The rotational
revivals can now be regenerated at a later time.  For example, in
Figs.~\ref{fig:control_revivals}(c) and (d), the revivals are regenerated using a
final control pulse (again identical to the first pump pulse) positioned at 40
ps and 70 ps, respectively.

\begin{figure}
    \centering
    \includegraphics[width=\columnwidth]{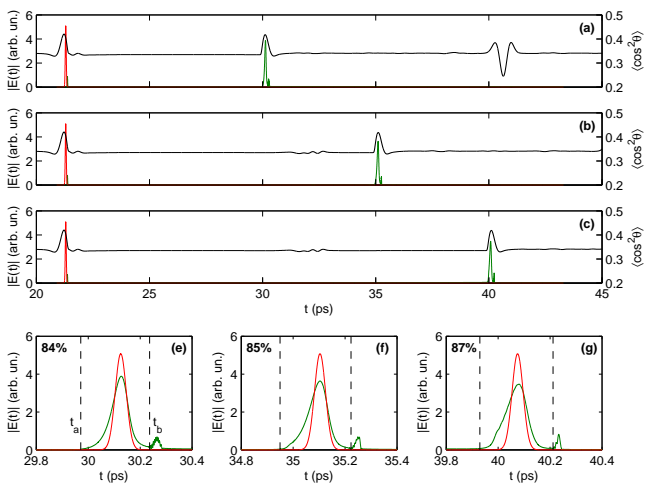}
	\caption{Examples of variable read times.  A signal pulse is stored in the memory
	at 21.3 ps (red pulse).  The memory is then read out at 30 ps (a), 35 ps (b), and 40 ps (c). 
	The re-emitted signal read out at these various times is shown in green.
	Panels (e-g) show close-ups of the emittied signal, along with a time-shifted copy of the
	intput signal for comparion.  The percentages in panels (e-g) show the efficiencies
	of the quantum memory for each case.  }
    \label{fig:control_read}
\end{figure}

We now demonstrate controllable read times for the memory scheme using this
control over the alignment revivals.  Fig.~\ref{fig:control_read} shows example
calculations where the read time has been set to three different values.
These calculations use a medium length of 4 cm and an atomic density
of $N_a = 1.6 \times 10^{16}$\,cm$^{-3}$.  All remaining parameters are
the same as used in Fig.\ref{fig:O2_memory}.
Consider first Figs.~\ref{fig:control_read}(a)-(c).  In these cases, the 50 fs signal
pulse (red) arrives at 21.3 ps and is stored in the memory using the falling edge of
concomitant revival.  This revival is then immediately attenuated with a
control pulse positioned at 21.45 ps.  Panels (a)-(c) show cases where the 
stored signal is then read out at three different times.  The read-out is
triggered by regenerating the revivals at times 30, 35, and 40 ps
respectively using a final control pulse.  
The re-emitted signal pulses are shown in green. In the plots, the re-emitted
pulses are seen to lag a bit (on the order of 0.1 to 0.2 ps) behind the
read-out revival.  This occurs because, once re-emitted, the signal pulse
propagates in a region of $n(t)$ that has a higher value than the pump pulse,
and hence the re-emitted signal starts to lag behind the alignment revivals
which are propagating with the speed of the pump.

Figures~\ref{fig:control_read}(e)-(g) show a closer
view of the emitted signal pulses (green), together with a time-shifted copy of
the initial signal pulse (red) for comparison.  The re-emitted signals are
seen to be broader in time as compare to the initial signal.  This is due
to the fact that $n(t)$ is not perfectly linear; the non-linear components
of $n(t)$ add dispersion to the propagation and effectively add chirp the signal.
Also seen in Fig.~\ref{fig:control_read}(e)-(g) is a small post-pulse, which
is a leakage process of the memory.  For these cases, we also calculate
the efficicency ${\cal E}$ of the memory by integrating over the intensities in
the main part of the signal 
\begin{equation}
	{\cal E} = \frac{ \int_{t_a}^{t_b} |E_{out}(t)|^2 dt }{ \int |E_{in}(t)|^2 dt }
\end{equation}
where $|E_{out}(t)|^2$ and $|E_{int}(t)|^2$  are the intensities of the output
(emitted) and input (stored) signal pulses.  The integration region over the
main part of the output pulse, $t_a$ to $t_b$, is labeled explicitely in
Fig.~\ref{fig:control_read}(e), while in 
Figs~\ref{fig:control_read}(f)-(g) we denoted this region by the vertical
dashed lines alone.  The resulting efficiencies are shown as percentages in
(e)-(g) and are between 84\% to 87\%.  Although
not seen clearly, part of the deviation from 100\%
efficiency is due to less than perfect efficiency in the write step.

These examples show a quantum photonic memory that stores a 50 fs pulse for
about 20 ps, giving a time-bandwidth product of approximately 20 ps / 0.05 ps =
400.  The largest storage times acheivable is limited by the $T_2$ time of the two-level atoms,
typically on the order of one nanosecond. For a 50 fs pulse, this leads to a
time-bandwidth product of order $~10^4$.


\subsection{Additional efficiency considerations}

\begin{figure}
    \centering
    \includegraphics[width=\columnwidth]{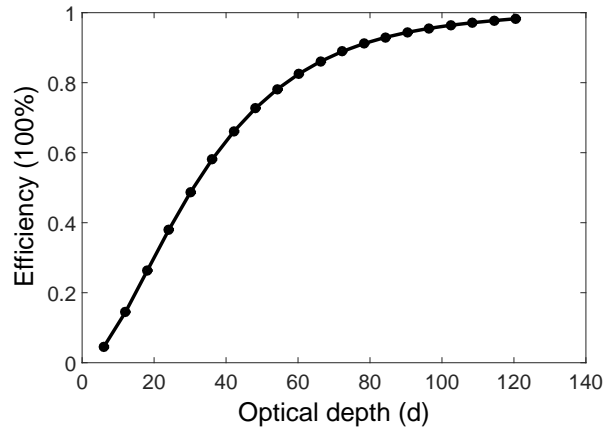}
	\caption{Total memory efficiency as a function of optical depth of the
medium, as defined in Eq.~\ref{eqn:opticaldepth}. High optical depths are
required as the scheme involves storing large bandwidth pulses in an atomic
transition with a narrow linewidth.}
    \label{fig:efficiency}
\end{figure}

The efficiency of the quantum memory is plotted
as a function of optical depth in Fig.~\ref{fig:efficiency}. The optical depth
of the system is modified due to the effective inhomogeneous broadening of the
medium. If we assume the refractive index varies linearly in the region where
the pulse is absorbed and emitted, \textit{i.e.} $n(t) = n_0 + \dot{n}t$, then
the optical depth is given by~\cite{clark2012photonic}
\begin{equation}
d = \frac{2\pi N_a \mu_{ba}^2}{\omega_0}\frac{n_0}{\dot{n}} \approx \frac{2\pi
N_a\mu_{ba}^2}{N_m\omega_0\Delta \alpha
\frac{d}{dt}\langle\cos^2{\theta}\rangle_T(t)}
\label{eqn:opticaldepth}
\end{equation}
in the regime where $N_m \ll 1$.  A substantial optical depth is required for
high efficiency as the atomic ensemble is absorbing large-bandwidth pulses with
a very narrow linewidth. Such optical depths have been achieved in hollow-core
fibers containing warm atomic
vapors~\cite{2015arXiv150904972K,sprague2013efficient}.  With this in mind, it
is reasonable to use the D$_1$ transition of an ensemble of $^{87}$Rb atoms in
a warm atomic vapor state for the gaseous atoms. The lifetime of this
transition is on the order of nanoseconds and as such can be neglected in
Eq.~\ref{eqn:bloch}. Furthermore, we also note that room temperature diatomic
molecules can be aligned in hollow-core fibers~\cite{sickmiller2009effects}.
Compared to the above simulations that model a high-pressure cell regime,
implementing the memory in a hollow-core fiber would force a lower molecular
density $N_m$ and hence would require a longer propagation length on the order of
1\,m.  Thus the proposed scheme could be achieved in either a high-pressure gas
cell a hollow-core fiber or containing a mix of a diatomic gas and a warm
atomic vapor.

\section{Conclusion}
\label{sec:conclusion}
In summary, we explore the possibility of implementing a GEM-type quantum
memory for ultrafast photons using molecular alignment to generate a
time-dependent refractive index~\cite{clark2012photonic}: A linear refractive
index is used to scan the frequency distribution of a broadband photon over the
transition frequency of an ensemble of two-level atoms such that entire photon
is absorbed. Later in time, the rotating molecules generate the opposite
gradient in the refractive index and the photon is re-emitted. We numerically
demonstrate that the proposed scheme can store a weak 50-fs pulse for 400 times
its duration in a gas mixture of $^{87}$Rb (warm atomic) vapor and CO$_2$
molecules at room temperature. This is achieved without the use of an external
static field as required in previous GEM schemes. By exploiting the revival
dynamics of the rotational wave packet generated in the alignment process along
with additional control pulses, we envisage that the lifetime of such a memory
can be extended to $T_2$ timescales of $^{87}$Rb, leading to time-bandwidth
products of order $10^{4}$. 

The proposed scheme explores one possible way of
implementing an optical quantum memory in a medium with an ultrafast
time-dependent refractive index. Inducing refractive index changes with
ultrafast response times is also achievable via other mechanisms, such as the
non-linear response of the electronic polarization of the
medium~\cite{boyd2003nonlinear} and in particular cross-phase
modulation~\cite{spanner2003controlled}.

\section*{Acknowledgments}
The authors thank Euan Joly-Smith, Davor Curic, Rune Lausten, and, Paul Hockett
for stimulating discussions.

\end{document}